\documentclass{PoS}

\usepackage[numbers]{natbib}
\usepackage{amssymb}

\makeatletter
\g@addto@macro\normalsize{%
  \setlength\abovedisplayskip{5pt}
  \setlength\belowdisplayskip{5pt}
  \setlength\abovedisplayshortskip{0pt}
  \setlength\belowdisplayshortskip{5pt}
}
\makeatother

\title{The Multi-Mission Maximum Likelihood framework (3ML)}

\ShortTitle{3ML}

\author{\speaker{Giacomo Vianello}\\
        Stanford\\
        E-mail: \email{giacomov@stanford.edu}}

\author{Robert J. Lauer\\
        University of New Mexico\\
        E-mail: \email{rlauer@phys.unm.edu}}

\author{P.~Younk (LANL), L.~Tibaldo (KIPAC/SLAC), J.~M.~Burgess (KTH Royal Institute of Technology), H.~Ayala (MTU), P.~Harding (LANL), M.~Hui (MTU), N.~Omodei (Stanford), H.~Zhou (MTU)}

\abstract{Astrophysical sources are now observed by many different instruments at different wavelengths, from radio to high-energy gamma-rays, with an unprecedented quality. Putting all these data together to form a coherent view, however, is a very difficult task. Each instrument has its own data format, software and analysis procedure, which are difficult to combine. It is for example very challenging to perform a broadband fit of the energy spectrum of the source. The Multi-Mission Maximum Likelihood framework (3ML) aims to solve this issue, providing a common framework which allows for a coherent modeling of sources using all the available data, independent of their origin. At the same time, thanks to its architecture based on plug-ins, 3ML uses the existing official software of each instrument for the corresponding data in a way which is transparent to the user. 3ML is based on the likelihood formalism, in which a model summarizing our knowledge about a particular region of the sky is convolved with the instrument response and compared to the corresponding data. The user can choose between a frequentist analysis, and a Bayesian analysis. In the former, parameters of the model are optimized in order to obtain the best match to the data (i.e., the maximum of the likelihood). In the latter, the priors specified by the user are used to build the posterior distribution, which is then sampled with Markov Chain Monte Carlo or Multinest. Our implementation of this idea is very flexible, allowing the study of point sources as well as extended sources with arbitrary spectra. We will review the problem we aim to solve, the 3ML concepts and its innovative potential.}

\FullConference{The 34th International Cosmic Ray Conference,\\
		30 July- 6 August, 2015\\
		The Hague, The Netherlands}

\begin{document}

\section{Introduction}

Multi-wavelength data contain fundamental physical information about the nature of astrophysical sources, the emission mechanisms, and information about the Universe at large (cosmology), representing a big opportunity for the advancement of astrophysics. This is widely recognized, as also reported in the NASA Decadal Survey \citep{astrophysics2011panel}. However, instruments at different wavelengths are based on different technologies, requiring different handling, software, skills and resources. A joint analysis is hence difficult and sometimes impossible. The Multi-Mission Maximum Likelihood framework (3ML) aims to solve this problem. In order to understand its innovative approach, we need to look at the observation process and how the problem we just presented has been solved in the past. 

\begin{figure*}[t!]
\centering
\includegraphics[width=0.65\textwidth]{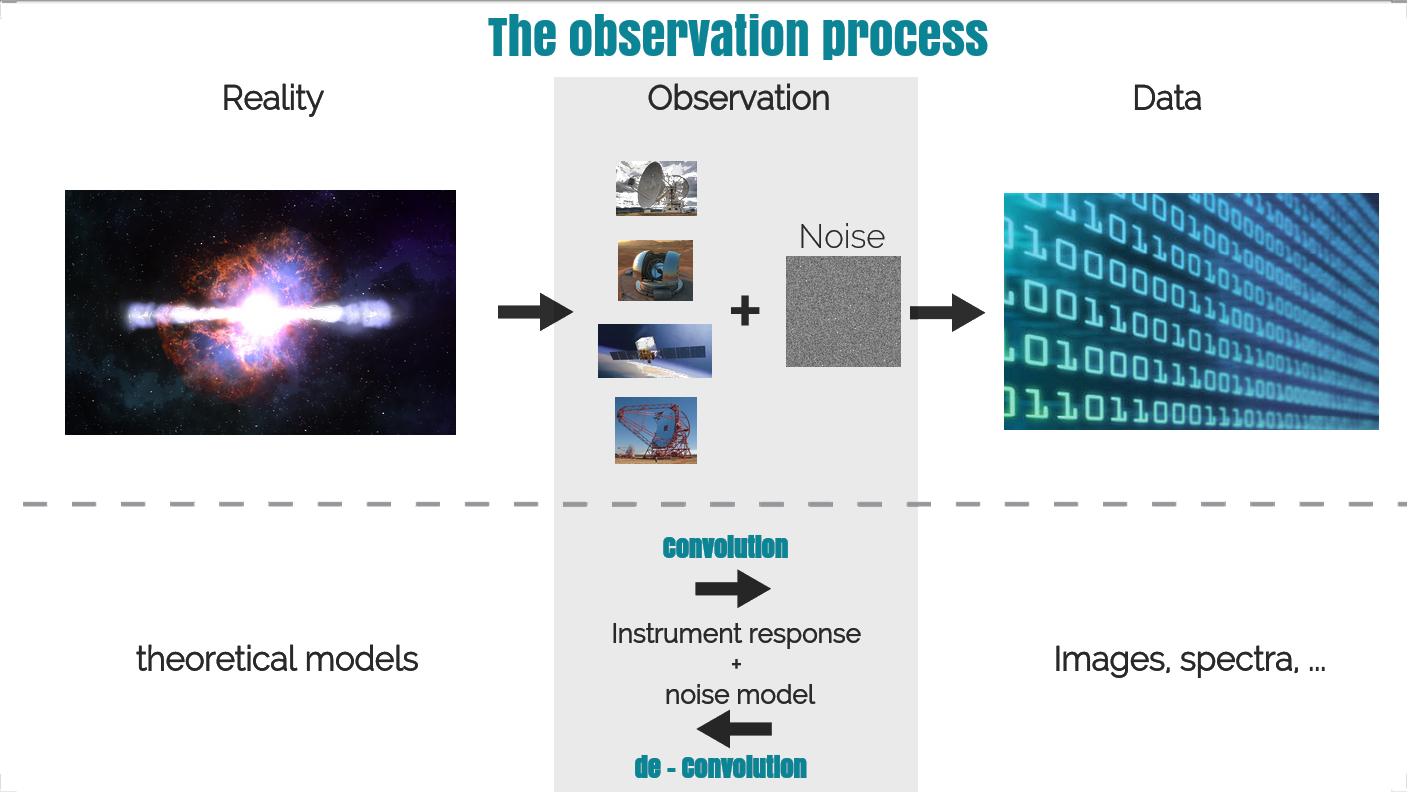}
\caption{The observation process.}
\label{fig:observation}
\end{figure*}

The process of observation is represented in Fig.~\ref{fig:observation}. The sky (\textit{reality}) is observed by a telescope, which collects photons, measure their properties (direction, energy etc.) and save them in a data set. In a certain sense, the observation process brings the signal from the reality domain to the data domain. It also necessarily adds noise to the data. We call \textit{irreducible} the noise inherent to the observation, such as Poisson noise when counting photons. We instead call \textit{background} the noise coming from the apparatus or the environment (detector noise, light pollution, or sources other than the one under study). In order to understand an astrophysical source, we need to connect theoretical models, which describe reality and live in that domain, to the data domain. Let us now formalize the problem. In the following upper-case and lower-case letters indicate respectively quantities in the reality and in the data domain. We also express area in cm$^{2}$, energy in keV, time in seconds and solid angle in steradians. Be $S \equiv S(E,\vec{P})$ the \textit{true} differential flux of photons coming from the sky position $\vec{P}$ at the energy $E$, in units of photons cm$^{-2}$ s$^{-1}$ keV$^{-1}$ sr$^{-1}$. An instrument, which is the bridge between reality and data, is characterized by its \textit{response} $R \equiv R(e,E,\vec{p},\vec{P})$, which is a function of the \textit{true} energy $E$ and position $\vec{P}$, and of the \textit{reconstructed} energy $e$ and position $\vec{p}$. We call \textit{energy dispersion} the phenomenon for which $R(e,E, \cdot, \cdot) > 0$ for $e \neq E$, while the phenomenon for which $R(\cdot, \cdot, \vec{p}, \vec{P}) > 0$ for $\vec{p} \neq \vec{P}$ is related to the optical properties of the instrument, summarized by the so-called \textit{Point Spread Function} (PSF). We assume that it is possible to factorize $R$ as $R = A(E,\vec{P})~PSF(E, \vec{p}, \vec{P})~D(e,E)$, where: A is the geometric cross section of the detector visible by incident radiation coming from the direction $\vec{P}$ multiplied by the efficiency at energy $E$ (\textit{effective area}, in cm$^{2}$); PSF is the Point Spread Function in sr$^{-1}$ normalized to 1, i.e., $\int PSF(E,\vec{p},\vec{P})~d\vec{p} = 1$; and D is the energy dispersion in keV$^{-1}$ normalized to 1, i.e., $\int D(e,E)~de = 1$. For simplicity we assume that $R$ does not vary during our observation. The differential rate of photons measured by our instrument at position $\vec{p}$ and energy $e$ is:
\begin{equation}
n(\vec{p}, e)~= \int~\int~\left[S(E,\vec{P})~A(E,\vec{P})~PSF(E,\vec{P},\vec{p})~D(E,e)\right]~dE~d\vec{P}~~~~~~[\mbox{photons s$^{-1}$  keV$^{-1}$ sr$^{-1}$}].
\label{eq:one}
\end{equation}
Typically an instrument measures radiation in $i=1..m$ spatial bins and $j=1..k$ energy bins of finite size. For example, an instrument with a CCD camera has spatial bins corresponding to the pixels in the camera. The number of photons detected in the $i,j$ bin in the time interval $\Delta t$ \textit{in the absence of noise} is:
\begin{equation}
o_{ij}~= \Delta t~\int_{i}~\int_{e_{j,1}}^{e_{j,2}}~n(\vec{p},e)~d\vec{p}~de~~~~~~[\mbox{photons}],
\label{eq:two}
\end{equation}
where the first integral is performed on the space covered by the $i$-th spatial pixel, and $e_{j,1}$ and $e_{j,2}$ are the boundaries of the $j$-th energy bin. What we actually measure is a random variable:
\begin{equation}
\widehat{o}_{ij} = P(o_{ij}, ~..) ~~~~~~[\mbox{photons}],
\label{eq:prob}
\end{equation}
where $P(o_{i,j}, ~..)$ is a probability distribution describing our \textit{irreducible} noise. It generally depends on $o_{ij}$ and possibly on other variables, such as the residual background. For example, if we are counting photons, $P$ is a Poisson distribution. We can now state the problem we aim to solve: we have a set $d$ of measurements $\widehat{o}_{ij}$ (data), we know\footnote{Usually we know R up to a certain degree. Our uncertainty about the response translates in systematic uncertainties, which are outside the scope of this paper and will be neglected here.}$R$ and the noise distribution $P$, and we want to find out as much as we can about $S$. This can be accomplished in two ways: i) \textit{bringing a model for S to the data domain} by convolving it with $R$, and then comparing this \textit{folded} model with the data through for example a \textit{likelihood analysis} based on the appropriate noise distribution $P$, ii) on the converse \textit{bringing the signal back to the reality domain} by de-convolving the data by $R$, and comparing such \textit{unfolded} data with a theoretical model. This second strategy is the most common for multi-wavelength data, and we therefore analyze it first in the next section.

\section{Multi-wavelength analysis of a source}

For simplicity we assume in this section that our instruments are pointed on a region of the sky with only one point source (for example an AGN or a star) at the position $\vec{P}_{\star}$, and we neglect any residual background\footnote{Although this is not realistic, a proper treatment of the background would complicate the notation, and it is not needed for the purposes of this paper.}. Then $S \propto \delta_{k}(\vec{P},\vec{P}_{\star})$, where $\delta_{k}$ is Kronecker's delta. We drop the $\vec{P}$ and $S$ is now the \textit{true} spectral energy distribution function for the point source.

\subsection{Bringing the signal back to the reality domain: Spectral Energy Distribution}

The problem of analyzing multi-wavelength data has been classically approached by using the de-convolution process. For point sources this means building the so-called Spectral Energy Distribution, which is a plot of the differential energy flux measured at different energies (or wavelengths). Such SED is then studied by fitting theoretical models to it. 
Let us consider one energy bin centered on $\tilde{e}$ with a very narrow band pass $\delta e$ and no energy dispersion. In this case $D(e,E) = \delta_{k}(e,E)$, i.e., for every photon the measured energy is equal to the true energy. Hence, $R = A(e,\vec{P}_{\star})~PSF(e,\vec{p},\vec{P}_{\star})$, with $R > 0$ if and only if $\tilde{e} - \delta e/2 < e < \tilde{e} + \delta e/2$. Then, equation \ref{eq:one} becomes:
\begin{equation}
n(\vec{p}, e) = S(e)~A(e,\vec{P}_{\star})~PSF(e, \vec{p},\vec{P}_{\star}).
\end{equation}
Let us also use only one spatial bin covering the entire PSF of the instrument at all energies. We approximate eq.\ref{eq:two} as:
\begin{eqnarray}
o & \simeq & \Delta t~\int_{1}~S(\tilde{e})~ A(\tilde{e}, \vec{P}_{\star})~PSF(\tilde{e}, \vec{p},\vec{P}_{\star})~ \delta e~d\vec{p} \\
& = & \Delta t~S(\tilde{e})~\delta e~A(\tilde{e}, \vec{P}_{\star})~\int_{1}~PSF(\tilde{e},\vec{p},\vec{P}_{\star})~d\vec{p} \\
& = & \Delta t~S(\tilde{e})~\delta e~A(\tilde{e}).
\end{eqnarray}
We then equate $o$ to our measured counts $\widehat{o}$ to build a measurement $\widehat{S}$ for $S(\tilde{e})$ as:
\begin{equation}
\widehat{S}(\tilde{e}) = \frac{\widehat{o}}{\delta e~\Delta t~A(\tilde{e})}~~~~~~\mbox{[photons cm$^{-2}$ s$^{-1}$ keV$^{-1}$]}.
\label{eq:easysed}
\end{equation}
A measurement for the \textit{energy} flux $\widehat{F}(\tilde{e})$ is obtained by multiplying $\widehat{S}$ by the \textit{typical} energy of the photons in the energy band of interest and by the bandwidth $\delta e$. Since $\delta e$ is small, the typical energy is $\tilde{e}$ and\footnote{The following quantity can be converted to the more typical unit of erg cm$^{-2}$ s$^{-1}$ by multiplying it by $1.602 \times 10^{-9}$ erg keV$^{-1}$.}:
\begin{equation}
\widehat{F}(\tilde{e}) = \tilde{e}~\delta e~\widehat{S}(\tilde{e}) = \frac{\tilde{e}~\widehat{o}}{\Delta t~A(\tilde{e})}~~~~~~\mbox{[keV cm$^{-2}$ s$^{-1}$]}.
\label{eq:sed}
\end{equation}
If the data from all instruments can be divided in bins small enough for these approximations to hold, we can build the SED by repeating this procedure for all bins. Otherwise, eq.\ref{eq:one} cannot be simplified. This can happen for example if the energy dispersion for our instrument is large, or if the source is not bright enough to divide the signal in many bins. In order to proceed we are forced to \textit{assume} a mathematical form for the function $S(E)$. Let us write $S(E) = k~\mathbb{S}(E)$, where we have singled out the normalization $k$ and the \textit{shape} $\mathbb{S}(E)$. For example, for a power-law, $\mathbb{S}(E) = E^{-\alpha}$. One way to choose such a function is to fit the entire dataset for each instrument, finding a shape providing a good fit\footnote{Note that using directly such a best fit to get a measurement for the differential flux at different energies is a bad idea, because it would make all the points obtained this way not-independent and their errors artificially small. This would make it very difficult to use such points when modeling the entire SED, since most fitting procedures such as $\chi^{2}$ minimization assume independent data points.}. Then:
\begin{equation}
n(\vec{p},e) = k~\int~\mathbb{S}(E)~[A(E,\vec{P}_{\star})~PSF(E,\vec{p}, \vec{P}_{\star})~D(e,E)]~dE
\end{equation}
and (assuming again one spatial pixel covering the whole PSF):
\begin{equation}
o ~=~\Delta t~k~\int_{1}~\int_{e_{1}}^{e_{2}}~\int~\mathbb{S}(E)~[A(E,\vec{P}_{\star})~PSF(E,\vec{p},\vec{P}_{\star})~D(e,E)]~dE~de~d\vec{p}.
\label{eq:hardsed}
\end{equation}
$\mathbb{S}$ acts as a weighting function for the response. We are implicitly assuming that $\mathbb{S}$ is valid over the whole range in \textit{true} energy where $D(e,E) >0$, with $e_{j,1} < e < e_{j,2}$. We could now determine $k$ and the parameters of $\mathbb{S}$ by fitting this expression to our measured $\widehat{o}$. Unfortunately we have only one data point. If $\mathbb{S}=E^{-\alpha}$, for example, we have to fix $\alpha$, using the one obtained from a fit of the whole dataset for our instrument. We determine then a measurement $\widehat{k}$ and its error. The average energy for the photons in our energy bin is:
\begin{equation}
\langle e \rangle = \frac{~\int_{e_{1}}^{e_{2}}~E~\mathbb{S}(E)~dE}{ \int_{e_{1}}^{e^{2}}~\mathbb{S}(E)~dE}
\end{equation}
and finally we can take:
\begin{equation}
\widehat{F}( \langle e \rangle ) = \langle e \rangle ~(e_{2}-e_{1})~\widehat{k}~\mathbb{S}( \langle e \rangle )
\end{equation}
as our SED point.

It is clear then that the procedure is cumbersome, and that our results are \textit{model-dependent}, i.e.,  depend on the shape $\mathbb{S}$ we assumed. This dependence is strong for large energy bins and for instruments with large energy dispersion effects. A poor choice can bias the results introducing a systematic error in the measurement of the SED. The problem is particularly troubling for faint sources. First, we cannot divide the data too much because we need a good signal-to-noise in each bin. Moreover, suppose that our data for one particular instrument are not good enough to distinguish between a shape $\mathbb{S} = E^{-\alpha}$ and another one $\mathbb{S}^{\prime}=E^{-\alpha}~e^{-E_{c}}$, i.e., $\mathbb{S}$ and $\mathbb{S}^{\prime}$ give a comparably-good fit to the data. Which function shall we use as weighting? The situation is even more confused. After having build the SED for all the instruments - each instrument potentially with its own weighting function $\mathbb{S}$- we then fit the entire SED with our theoretical model, which might differ substantially from the weighting functions used. This is incoherent and prone to errors. Also, this procedure is difficult to generalize to extended sources. Luckily there is a better option for these cases, which we will analyze in the next section.

\subsection{Bringing the model to the data domain: forward folding}

Suppose that we have a good understanding of the physical processes happening in a source, and we are able to write a model for the spectrum of such source $S(E|\vec{\alpha})$, where $\vec{\alpha}$ is a vector of parameters whose values we want to constrain. We use eq.~\ref{eq:one} and eq.~\ref{eq:two} to obtain the predicted number of counts $o_{ij}$ in any given bin (\textit{forward-folding}). We can \textit{always} do this, no matter the size of the bins or the characteristics of the instrument. We rewrite the probability function $P$ introduced in eq.~\ref{eq:prob} as the probability $P(\widehat{o}_{ij}|o_{ij})$ of observing $\widehat{o}_{ij}$ photons given the expectation $o_{ij}$. For example, if $P$ is a Poisson distribution, then $
P(\widehat{o}_{ij}|o_{ij}) = \frac{o_{ij}^{\widehat{o}_{ij}}~e^{-o_{ij}}}{\widehat{o}_{ij}!}.
$
We do this for all our bins. Let $d$ be the collections of all the $\widehat{o}_{ij}$ (i.e., our data set), which are independent measurements. The joint probability of obtaining $d$ given S and $\vec{\alpha}$ is:
\begin{equation}
L(d | S, \vec{\alpha}) = \prod_{ij}~P(\widehat{o}_{ij}|o_{ij}).
\label{eq:likelihood}
\end{equation}
This is called \textit{likelihood} function, and expresses the probability of obtaining $d$ if $S(E,\vec{\alpha})$ is the \textit{true} model. We can now look for the set $\vec{\alpha}_{MLE}$ which maximizes the likelihood (\textit{Maximum Likelihood Estimation}, or MLE, \citet{bohm2010introduction}), which is our best fit to the data. Computationally it is more convenient to minimize the function $-\log{(L)}$ instead of maximizing $L$, but clearly $\vec{\alpha}_{MLE}$ which minimizes $-\log{(L)}$ is the same which maximizes $L$. We can build a likelihood function for any choice of $P$. For example, the usual $\chi^{2}$ function is (twice) the log-likelihood function when $P$ is a Gaussian distribution. Also, we can make a combined likelihood for datasets with different $P$ distributions by simply summing the corresponding log-likelihood functions. If we have competing models for the same source, or if we do not know which $S$ to use, we can explore different solutions and use the likelihood function for model selection (for example through the Likelihood Ratio Test). $L$ can also be used to estimate the \textit{goodness-of-fit} for our model. Unfortunately both these operations often require extensive Monte Carlo simulations. We can also combine $L$ with \textit{prior} knowledge and perform Bayesian inference, which also supplies alternative ways for model selection. The likelihood approach allows to study multiple sources at once, even extended sources, as long as we are able to formulate a function $S(\vec{P},E)$. Hence forward-folding combined with likelihood or Bayesian analysis is a powerful tool which is always applicable, even when a model-independent SED is impossible or when we are dealing with many or extended sources. 

\section{The Multi-Mission Maximum Likelihood framework (3ML)}

\begin{figure*}[tb]
\includegraphics[width=0.35\textwidth]{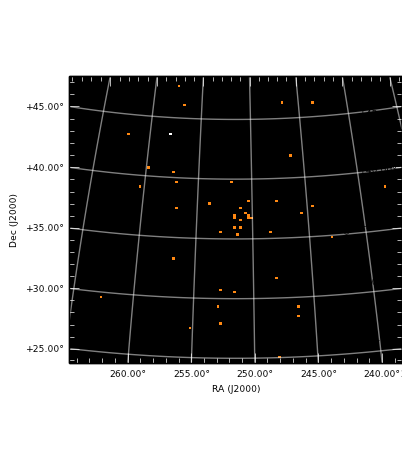}
\includegraphics[width=0.6\textwidth]{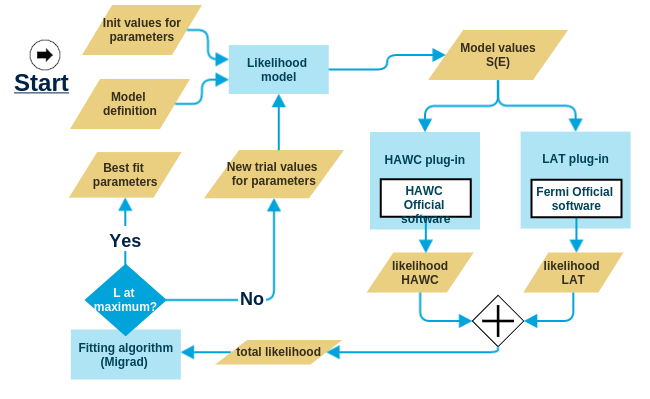}
\caption{Left panel: simulation of a faint point source with Fermi/LAT. Right panel: an example of a likelihood maximization with 3ML.}
  \label{fig:design}
\end{figure*}

The forward-folding approach described in the previous section has been already adopted in the X-ray community. Software such as XSPEC \citep{2014HEAD...1411504A}, ISIS \citep{2008PASP..120..821N} or Sherpa \citep{2011ascl.soft07005F} adopt such formalism. Unfortunately they handle only the OGIP format for the data\footnote{http://heasarc.gsfc.nasa.gov/docs/heasarc/ofwg/docs/spectra/ogip\_92\_007/ogip\_92\_007.html}, which has been designed for X-ray data. This format is not well suited for observatories where the spatial information on single events is important. For instance, let us consider the Fermi Large Area Telescope \cite{2012ApJS..203....4A}. LAT is a gamma-ray telescope, sensitive from 20 MeV to over 300 GeV, with imaging capabilities. Let us consider the case of a Gamma-Ray Burst, with a duration < few minutes. The typical background for the LAT is very low and a spatial PSF-like cluster of few events can constitute a solid detection on such a time scale. Accordingly, an appropriate likelihood function for LAT data must account for the spatial distribution of the events. To illustrate this point, we simulated a LAT observation of 100 s of a point source and a typical background (Fig.~\ref{fig:design}). There are 45 background counts and 15 source counts. The significance of the point source obtained with the ad-hoc LAT likelihood software, which accounts for the spatial structure of the data, is more than $7 \sigma$. If we ignore that spatial structure, the Poisson probability of obtaining $45 + 15 = 60$ counts or more when expecting $45$ is 2\%, corresponding to less than 2 $\sigma$. We can now see how transforming the data in OGIP format, which requires integrating over the spatial dimension, results in a big loss of sensitivity. A similar but even more extreme case is represented by the data from the High-Altitude Water Cerenkov experiment (HAWC), an array of water Cherenkov detectors instrumented with photomultipliers (PMTs), sensitive from 100 GeV to 100 TeV \cite{HAWChighlight}. On top of the problem with the spatial resolution, which is important for HAWC as well, the latter also has a poor energy resolution due to technological limitations. For this reason, its response $R$ is better formulated as a function which translates true energies for the incoming photon in number of hits recorded by the PMTs (nHits) instead of observed energy \citep[see][for details]{HAWCLiFF}. Encoding such a response in the OGIP format is impossible. Similar problems are present for other high-energy observatories, such as VERITAS, HESS and MAGIC. Moreover, it is of course impossible to study the morphology of an extended source such as a Supernova Remnant using OGIP. These problems motivated us to create the Multi-Mission Maximum Likelihood framework.

The need for flexibility described above led us to create a software architecture based on \textit{plug-ins}, each one handling the data from one instrument. Each operating telescope has already its own official software (OS), developed for that observatory only. A plugin is an interface between 3ML and the OS of one instrument: it receives from the framework the likelihood model $S(\vec{P},E|\vec{\alpha}_{s})$, where $\vec{\alpha_{s}}$ is a particular set of parameters. It then uses the OS to compute the value for the likelihood function  appropriate for that instrument for the set $\vec{\alpha}_{s}$. This architecture guarantees the best treatment of the data, avoids placing any constraint on how data should be formatted or treated, and minimizes the effort for including a new instrument capitalizing on the existing OS. To illustrate this point, let us consider a simple analysis for a single point source, using data from HAWC and Fermi/LAT. The analysis flow is shown in the right panel of Fig.~\ref{fig:design}: a model definition $S$ with its parameters and initial values $\vec{\alpha}_{s}$ are fed in 3ML. The model values $S(E,|\vec{\alpha}_{s})$ are passed to the plugins, which compute the likelihood values using the OSs of the two instruments. These values are then summed and passed to the fitting engine (MIGRAD in this case) which decides if the fit has converged. If not, new trial values $\vec{\alpha}_{s,new}$ are generated and the loop restarted. The key point here is that 3ML does not need to know \textit{how} the plug-ins uses the OSs to compute the respective likelihood. Hence, each OS is free to use the language, the data format, the analysis procedure and the likelihood formula which best fits its needs. As clear from the formalism presented in the introduction, the likelihood model $S$ can contain as many point sources as needed, and extended sources as well. For the first time, then, 3ML allows multi-wavelength modeling of extended sources. This is particularly interesting for combining instruments with wide field of view but comparably coarse spatial resolution with instruments with smaller FOV but higher spatial resolution, such as for example HAWC and VERITAS.
Our architecture allows also to naturally account for the so-called \textit{nuisance} parameters. For instance, let us consider again the Fermi/HAWC combined analysis. The background for LAT is made up of two components, the Galactic and the isotropic background component, which are known up to a normalization constant which must be fitted to the data. The background in HAWC is different and has its own normalization. These background normalizations are not of immediate interest for the study of the source, but must be taken into account in the analysis. Parameters like these are called \textit{nuisance} parameters. We divide $\vec{\alpha} = (\vec{\alpha}_{s}, \vec{\alpha}_{LAT}, \vec{\alpha}_{HAWC}$), where $\vec{\alpha}_{s}$ are the parameters for the source, while $\vec{\alpha}_{LAT}$ and $\vec{\alpha}_{HAWC}$ are parameters for the background in LAT and HAWC respectively. While $\vec{\alpha}_{s}$ plays a role in the likelihood for both instruments, $\vec{\alpha}_{LAT}$ does not play a role in the HAWC likelihood, nor $\vec{\alpha}_{HAWC}$ in the LAT likelihood. We can write a \textit{profile} likelihood \citep{bohm2010introduction} $\log{L_{tot}}$ which only depends on $\vec{\alpha}_{s}$, with:
\begin{equation}
-\log{L_{tot}(\vec{\alpha_{s}})} = \min_{\vec{\alpha_{LAT}}}{[-\log{L_{LAT}}(\vec{\alpha}_{s},\vec{\alpha}_{LAT})]} + \min_{\vec{\alpha_{HAWC}}}{[-\log{L_{HAWC}}(\vec{\alpha}_{s},\vec{\alpha}_{HAWC})]},
\end{equation}
where $\min_{x}{[f(\vec{x},\vec{y})]}$ denotes the minimum of $f$ with respect to $\vec{y}$ with $\vec{x}$ fixed. Practically, the fitting engine only works on the $\vec{\alpha}_{s}$. For each trial set $\widehat{\vec{\alpha}}_{s}$ the LAT plugin executes a \textit{inner} fit in which the LAT likelihood is maximized with respect to $\vec{\alpha}_{LAT}$ with $\widehat{\vec{\alpha}}_{s}$ fixed, and the HAWC plugin does the same for $\vec{\alpha}_{HAWC}$. In other words the fitting engine maximizes the \textit{profile} likelihood in which $\vec{\alpha}_{LAT}$ and $\vec{\alpha}_{HAWC}$ are \textit{profiled out} by the plugins. We can also deal with inter-calibration between instruments, introducing a normalization constant for each instrument as a nuisance parameter.

In conclusion, the architecture of 3ML provides the flexibility needed for multi-wavelength studies. It is even possible to implement a multi-messenger analysis, where data from neutrino or gravitational wave experiments are combined with electromagnetic data. As long as our model $S$ can make predictions about neutrino or GW fluxes, plugins for such instruments could provide the corresponding likelihood. 3ML is open-source\footnote{Code is available at threeml.stanford.edu}. It has been devised in the context of a coordination effort between the Fermi, VERITAS and HAWC collaboration. Its development, as well as plugins for Fermi/LAT, Fermi/GBM, Swift/XRT, Swift/BAT, Swift/UVOT, HAWC, VERITAS, HESS and optical telescopes with different filters are on-going. We will also provide a plugin for OSs based on GammaLib\footnote{http://gammalib.sourceforge.net/}, a toolbox for developing likelihood-based analysis for gamma-ray instruments which will likely be adopted by new experiments such as CTA.

\bibliography{bib}

\end{document}